# EVALUATING FUNCTION-AS-A-SERVICE (FAAS) FRAMEWORKS FOR THE ACCELERATOR CONTROL SYSTEM

A. Jaikar*, J. Diamond, A. Tiradani, B. Harrison
Fermi National Accelerator Laboratory, Batavia, IL, USA

## Abstract

The Fermilab Accelerator Complex is the largest national user facility in the Office of High Energy Physics (DOE/HEP) program which uses a linear accelerator (Linac), Booster rapid cycling synchrotron, Recycler storage ring, and Main Injector to produce proton beams. This particle accelerator produces high-volume, high-velocity telemetry and controls data that must be ingested, processed, and retained for operations and analysis. As particle accelerator control systems evolve in complexity and scale, the need for a responsive, scalable, and cost-effective computational infrastructure becomes increasingly critical. Function-as-a-Service (FaaS) offers an alternative to traditional monolithic architecture by enabling event-driven execution, automatic scaling, and fine-grained resource utilization. This paper explores the applicability of FaaS frameworks in the context of a modern particle accelerator control system, with the objective of evaluating their suitability for short-lived and triggered workloads.

## INTRODUCTION

The particle accelerator complex consists of thousands of devices like magnets, RF cavity, vacuum, power supply, and many other devices, which are coordinately controlled by the control system. This accelerator control system has been designed, developed, and maintained for around 3 decades. When the system was designed, there were many limitations, such as network latency, RAM capacity, CPU frequency, and network protocol as well. The system was designed with the ACNET (Accelerator Network) protocol [1], which is based on the User Datagram Protocol (UDP). Although it has performed very well over the years, modern technology landscapes demand higher efficiency, flexibility, and scalability, which the existing framework struggles to provide. This framework has been designed without the consideration of large storage, processing, and machine learning-based operations. Therefore, modernization is needed to support future operations of the complex.

In the accelerator complex, data is captured either at regular intervals or in response to specific events. Periodic data can be processed at regular intervals and event-based data can be processed at regular intervals or at the same time. To support both periodic and event-driven processing, the Function-as-a-Service (FaaS) has the ability to process the data with a scalable computing infrastructure. Kubernetes hosted FaaS platform helps the developer concentrate on developing a function which will process the data without managing the required infrastructure.

---
* amol@fnal.gov

## BACKGROUND

The particle accelerator produces heterogeneous high-volume data streams and operates under strict latency, reliability, and safety constraints. Therefore, data generated by these properties have the following characteristics:

- **High cardinality**: Slightly more than half million properties
- **Mixed frequency**: Different frequencies for different devices and their properties
- **Low latency**: Some application requires low latency data
- **Long retention**: months to years for trend analysis, reliability studies on historical data

Once the raw data has been ingested, it undergoes processing steps such as filtering, tagging, unit conversion, and other transformations. The FaaS platform offers event-based data processing that can be used for accelerator data processing. It also depends on Kubernetes for container orchestration, automatically handling scaling, networking, and resource allocation.

In this paper, we evaluate three Kubernetes based FaaS platforms; fission [2], Knative Serving/Eventing [3], and nuclio [4] with respect to their architecture, characteristics, and fit for the accelerator operations.

## ARCHITECTURAL OVERVIEWS

This section covers the architecture for all three FaaS frameworks.

### fission

fission is an open-source Function-as-a-Service (FaaS) framework built on Kubernetes, designed to provide a lightweight, event-driven execution environment. Instead of traditional server-based deployments, fission functions are short-lived, stateless, and triggered by events, making the framework highly suitable for cloud native microservices, event processing, and real-time data pipelines. Fission consists of multiple components that build up the architecture.

**Controller** The controller is the component which accepts requests from clients. All Custom Resource Definitions (CRDs) of fission, which are defined in Kubernetes, are used for Create, Read, Update, and Delete (CRUD) operations. A CRD is a powerful feature in Kubernetes that allows you to extend the Kubernetes API and define your own custom resource types. The controller uses CRUD APIs for functions, triggers, and environments. A cluster-wide admin permission is required for the controller to access all fission CRDs from all namespaces.

**Router** The router acts as a bridge between triggers and functions. It accepts HTTP requests and forwards to the respective function if there is one. If there is no running service for a function, it requests one from the executor. It is a stateless service that can scale depending on the requirement. Figure 1 shows the request flow. The first step shows that the request comes from the client to the router. In the second step, the router checks with the executor if the function is present to get the request. The third step is to obtain the response from the executor. Finally, the fourth step is to forward the incoming request to the appropriate function pod.

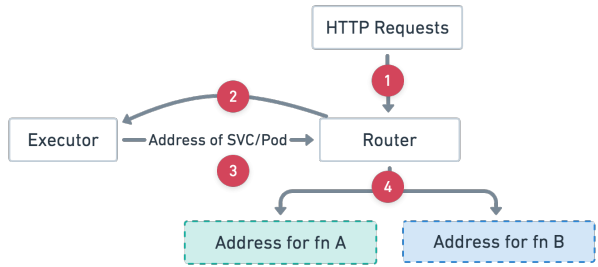

Figure 1: fission router [2].

**Executor** The executor is responsible for the execution of the requests coming from the router, which means spin up function pods for functions with the help of CRDs. It has two types of executor to spin up function pods, which have different strategies to launch, specialize, and manage pod(s). The first is PoolManager, and the second is NewDeployment. Figure 2 shows the architecture.

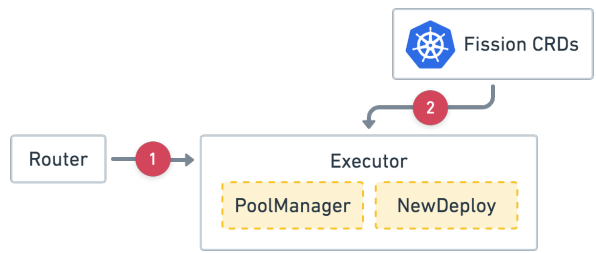

Figure 2: fission executor [2].

PoolManager manages pools of generic containers and function containers. The pool size of initial warm containers can be configured according to user needs. It pulls the generic pod and configures it according to the function. Once there are not enough requests for that function, then the pod is cleaned up to be used for another function. PoolManager is great for functions that are short-lived and require a short cold start time, which also depends on the function size.

NewDeployment creates Kubernetes deployment with Horizontal Pod Autoscaler(HPA) for function execution and makes it suitable for functions that handle massive traffic. It will scale the replicas for the function as per configuration. If there are no requests for a certain duration, then the idle pods are cleaned up. This approach not only increases the cold start time of a function, but also makes NewDeploy suitable for functions designed to serve massive traffic. If a function requires a minimum cold start time, then a minscale greater than zero can be set, which means that one pod is running even though there is no function call.

**Function pod** It contains two containers; one is the fetcher, and the other is the function environment. The fetcher is responsible for getting the function package from StorageSvc, which is a repository for source and deployment archives. The function environment pod is responsible for the execution of the function. It must have a HTTP server and loader for the function.

These are the main components of the fission framework. There are other components as well, but these are very important for understanding fission workflow.

*Knative*

Knative is an open source FaaS framework that extends Kubernetes to provide a platform for building, deploying, and managing modern serverless workloads. Knative's modular architecture has two primary components: Knative Serving and Knative Eventing. Knative serving has responsibility for request handling, revisioning, traffic management, and autoscaling. Knative eventing manages cloud event-based routing, brokers, channels, and sources.

Figure 3 shows the Knative-serving architecture. The activator accepts the incoming requests and keeps them in the queue to be served. Then, it communicates with Autoscaler to bring the service back if it is not present. The Autoscaler is responsible for scaling the service based on incoming requests, metrics, and service configuration. The controller is responsible for the Knative resources within a cluster. It not only manages the objects, but also the life-cycle of dependent resources as well. Queue-Proxy is a side-car container to collect metrics.

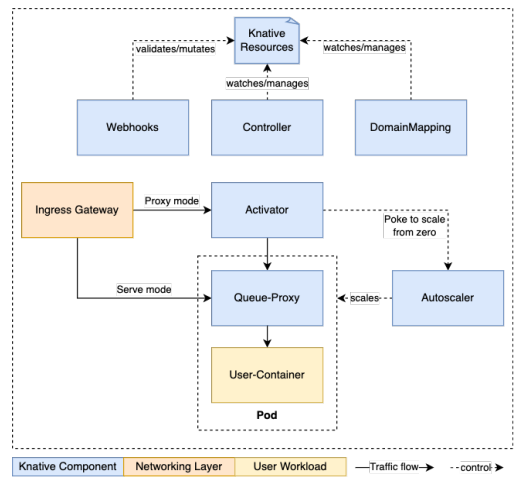

Figure 3: Serving architecture [3].

Knative eventing is used to enable an event-driven architecture with an application. Its components are loosely coupled and use HTTP POST requests to send and receive events. A publisher can publish a message without a consumer, and a consumer consumes the message from the intermittent broker. The event mesh is designed to distribute events from senders to recipients. It has the capability to design and execute very complicated workflows, which is suitable for an enterprise environment.

Knative deployment can have Knative serving and eventing together or only serving as well. Eventing is required for the management of external events.

### nuclio

nuclio is also an open source framework for serverless workload, which is designed for high performance data processing. Its architecture consists of a processor-based model, a highly optimized engine, and a flexible event-driven design. Each function is executed in the function processor. The function invocation has two elements; first one is an event object and the other is a context object. The event object contains data and metadata about the event object, which helps to decouple from event generation. Figure 4 shows the function processor architecture, which has event listeners, runtime engine, data bindings, and control framework.

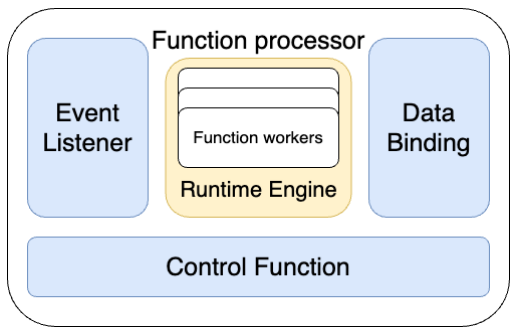

Figure 4: Function processor architecture [4].

**Event listeners**  The event listener's responsibility is to listen on sockets and message queues or fetch the events from external sources. The event listeners also guarantee exactly-once or at least once event execution with fault tolerance.

**Runtime engine**  The runtime engine initializes the function environment and feeds the event data to the function worker. Once the execution is completed, it returns the response to the event source. It can have multiple parallel workers for each function to enable non-blocking operation. Each event can have an EventProcessor that fetches the event. This newly created EventProcessor handles the event and forwards it to its runtime environment like Python, Go, Java, etc. The EventProcessor supports two modes; Synchronous, where events within a single worker are processed sequentially, and Asynchronous mode, where events within a single worker are processed concurrently.

**Data binding**  The runtime engine initializes data service connections according to type. Data binding eliminates the need to manage connections and credentials. Data bindings can also handle aspects of data prefetching, caching, and microbatching, to reduce execution latency and increase I/O performance. These features will enable real-time performance.

**Control framework**  The controller framework is responsible for controlling different components of the processing module(processor). It also provides logging for the processor and functions. Platform-specific processor configuration is done through a processor.yaml file in the working directory.

All these components play an important role in the management of nuclio framework.

## COMPARATIVE ANALYSIS

In the previous section, we have seen the architecture for each FaaS framework, which helps to understand the overall functionality. This section will give some of the differences based on analysis with consideration of accelerator complex data.

### Architecture

Fission tightly couples functions and environments together and executes them on the basis of PoolManager and the NewDeploy executor. Moreover, it decouples the build and execution pipeline. It also keeps some containers running to reduce the cold start time.

Knative's serving and eventing modules have different responsibilities. Serving provides an immutable deployment of the function with routing and scale-to-zero features. However, eventing takes care of complex event meshes. Therefore, Knative architecture is more suitable for enterprise-level modular based execution of complex workflow with multiple events.

nuclio is developed with the consideration of the high-performance worker model. The controller has the responsibility to run these optimized containers to respond to incoming triggers. It also supports batching and shared memory feature to reduce response time.

### Integration with Kafka

All three FaaS frameworks support Kafka integration with different approaches. Fission has kafka trigger built into its event system which can have one-to-one mappings between a topic and function. It also supports scalability with Kubernetes HPA feature.

Knative eventing supports triggering model with Kafka. The message pulled from Kafka topic will be converted into a cloud event which will be delivered to the Knative service module. This cloud event can be used for event meshing which can support filtering, routing, and fan-out to multiple triggers. Scaling of these environments will be handled by the Knative serving module.

The nuclio function connects directly to a Kafka broker. It gives the capability to process stream of messages quickly, but restricts the flexibility. Ref. [5] covers the performance with Kafka integration.

*API and Developer Ergonomics*

Fission works on REST api for functions, environments, packages, and triggers. It also supports a command line interface (CLI) to communicate with a fission framework. The event payload and CLI output is in JSON format by default.

Knative supports the YAML/JSON format for the configuration. It supports CLI with exposing HTTP endpoints for functions/services. The event payload is in cloud events, which has many options for integration. This feature makes it very flexible but difficult to adopt because of complexity and the event meshing workflow.

nuclio exposes functions through HTTP, CLI and simple user interface (UI). The event payload can be flexible such as JSON, binary, gRPC and many more.

*Cold Start and Throughput*

As the fission supports two execution models, PoolManager and NewDeploy, both have different cold start latency. PoolManager keeps warm pods active to reduce cold-start latency. However, NewDeploy has a higher cold-start latency due to pod creation on demand. Fission's router also introduces some latency for both execution models. Fission's latency and throughput have been covered in [6]. In addition, Knative's and nuclio's latency and throughput are covered in [7].

Knative has a slightly higher cold-start latency because of complexity and event mesh. However, scale-to-one instead of scale-to-zero reduces the cold-start impact.

nuclio is designed for low latency and high throughput. However, it requires more memory to run the function.

*Autoscaling and Security*

Fission takes advantage of Kubernetes HPA with the executor and provides scaling at the function level. However, it does not support request-based autoscaling like Knative. Knative also supports Knative Pod Autoscaler, making it best suitable for event-driven microservices. nuclio also has function-level autoscaling like fission but is optimized for real-time workload.

Regarding security, all three frameworks support the Kubernetes Role Based Access Control (RBAC) mechanism. However, Knative has more capability because of cloud events. It can integrate with multiple secret management platforms.

In addition to these features, [8] covers other parts of fission, Knative, and nuclio such as licensing, integration, GitHub metrics, and supporting language as well.

## CONCLUSION

With consideration of supporting data processing for accelerated data, it is very important to identify the FaaS platform which can satisfy most of the requirements. Fission provides a simple FaaS platform, which is easy to adopt and has low cold-start time as well. However, Knative supports event meshing with complex workflows, which is more suitable for enterprises. Fission and Knative also support decoupled integration with Kafka, which is not supported in nuclio.

Regarding API interfaces, all three frameworks support HTTP / REST. However, Knative has more integration options because of cloud events. On the other side, Knative has more cold start latency than fission and nuclio.

With the consideration of accelerator data, cold start latency, trigger support, and security, it is concluded that fission has sufficient features to support the data pipeline for the accelerator data.

## ACKNOWLEDGEMENTS

This manuscript has been authored by FermiForward Discovery Group, LLC under Contract No. 89243024CSC000002 with the U.S. Department of Energy, Office of Science, Office of High Energy Physics.

## REFERENCES

[1] F. J. Nagy, "The Fermilab accelerator controls networking system", *Nucl. Instrum. Methods Phys. Res., Sect. A*, vol. 247, no. 1, pp. 208–214, Jun. 1986.
doi:10.1016/0168-9002(86)90564-4

[2] fission homepage, https://fission.io/.

[3] Knative homepage, https://knative.dev/docs/.

[4] nuclio homepage, https://nuclio.io/.

[5] D. Balla, M. Maliosz, and C. Simon, "Performance Evaluation of Asynchronous FaaS", in *Proc. CLOUD 2021*, Chicago, IL, USA, Sep. 2021, pp. 147–156.
doi:10.1109/cloud53861.2021.00028

[6] G. I. Hegyi, M. Maliosz, and C. Simon, "Performance Measurements of Function as a Service Platforms", in *Proc. ISCSIC 2019*, New York, NY, USA, Sep. 2019, p. 43.
doi:10.1145/3386164.3390516

[7] J. Li, S.G. Kulkarni, K.K. Ramakrishnan, D. Li, "Analyzing open-source serverless platforms: Characteristics and performance", arXiv:2106.03601 [cs.DC], 2021.
doi:10.48550/arXiv.2106.03601

[8] V. Yussupov, J. Soldani, U. Breitenbücher, A. Brogi, and F. Leymann, "FaaSten your decisions: A classification framework and technology review of function-as-a-Service platforms", *J. Syst. Software*, vol. 175, p. 110906, May 2021.
doi:10.1016/j.jss.2021.110906